\documentclass[aps,prb,twocolumn,superscriptaddress,amsmath,amssymb,floatfix,nofootinbib]{revtex4-2}
\usepackage{graphicx}
\usepackage[dvipsnames]{xcolor}
\usepackage{ulem}
\usepackage{dsfont}

\graphicspath{{figures/}}

\begin{document}

\title{Rare-earth spin textures and a route to electronic inhomogeneity in $Pr_2Ir_2O_7$}

\author{Kyle G.\ Sherman}
\affiliation{Department of Physics, Applied Physics and Astronomy, Binghamton University, Binghamton, NY 13902}
\author{Michael J.\ Lawler}
 \email{mlawler@binghamton.edu}
\affiliation{Department of Physics, Applied Physics and Astronomy, Binghamton University, Binghamton, NY 13902}
\affiliation{Department of Physics, Cornell University, Ithaca, NY 14853}

\begin{abstract}
The pyrochlore iridate $Pr_2Ir_2O_7$ remains metallic at low temperatures where its family members insulate. Recently,  scanning tunneling spectroscopy experiments have revealed an inhomogeneous mixture of Kondo-screened and Kondo-destroyed regions when its surface is produced by cleaving at room temperature while it remains uniform when cleaved at low temperatures. We ask whether the frustrated praseodymium spin texture seeds this inhomogeneity. To study this, we produce Monte-Carlo-sampled Pr spin-ice configurations, presumed to evolve slowly in time, and couple through a local Kondo exchange $J_K$ to an eight-band Hartree-Fock model of the Ir electrons.

We further find the charge gap of the all-in-all-out state common in the other family members closes smoothly with the coupling $J_K$, and the recently proposed monopole-rich ``jellyfish'' textures further suppress the insulating behavior by $\Delta J_{Kc}\approx 0.03\,t$. Additionally, scanning a small cluster across a large Pr surface yields a synthetic tunneling map that fractures into islands within a sea. Although our classical simulable theory omits the Kondo singlet, the spatial modulation of the Fermi-level density of states it produces is, through the exponential Doniach sensitivity of the Kondo temperature, sufficient to tip the local balance between screening and magnetic order; a frustration-driven route to the observed Kondo/Kondo-destroyed inhomogeneity.
\end{abstract}

\maketitle

\section{Introduction}

Broadly, the pyrochlore iridates undergo a metal-to-insulator(MI) transition at low temperature. This transition is concomitant with AIAO magnetic order. Praseodymium iridate alone does not exhibit this transition down to the lowest measured temperatures~\cite{Matsuhira2011,Tomiyasu2012,Witczak-Krempa2014,Nakatsuji2006}. It also has the most ideal oxygen-octahedra in the family due to its large rare-earth radius. An increased overlap with the oxygen orbitals may facilitate Ir electron mobility in this case and tends praseodymium iridate toward metallicity~\cite{Matsuhira2011,Witczak-Krempa2014,Nakatsuji2006}. This is the gradual tendency with rare-earth size and may not explain the sudden drop in $T_{MI}$ for $Nd_2Ir_2O_7$ and $Pr_2Ir_2O_7$. Pr's large ionic radius also means that it serves as a large scattering center for the Ir conduction electrons, and thus Kondo physics is also present. 

Bulk and surface probes reveal a frustrated metallic Kondo lattice whose low-temperature behavior is set by the praseodymium non-Kramers doublet~\cite{Nakatsuji2006,Machida2007,Machida2010,Balicas2011,Tokiwa2014}. However, the magnetic ground state remains highly sensitive to strain, stoichiometry, and local environment~\cite{Kondo2015,Cheng2017,MacLaughlin2009,MacLaughlin2015,Ohtsuki2019}.

Scanning tunneling spectroscopy reveals Kondo physics in $Pr_2Ir_2O_7$, but with the caveat that it is inhomogeneously mixed with Kondo-destroyed regions~\cite{Kavai2021}. The screened and destroyed regions interleave on the nanometer scale, with power-law pattern statistics the authors read as proximity to a critical point. This result is contested by the work of Song et al who ascribe the inhomogeneity to the cleaving process~\cite{Song2025}. Working at $0.3$~K on atomically resolved $(111)$ surfaces, they report a homogeneous Kondo-lattice resonance. Whether the phase separation is intrinsic or an artifact of surface preparation is thus open. 

Theoretical treatments have historically isolated the Ir sublattice to reproduce the family-wide metal-insulator transition~\cite{Moon2013,Kondo2015,Wan2011,Witczak-Krempa2012,Savary2014}. Conversely, approaches incorporating the Pr moments have explored various $f$-$d$ exchange mechanisms~\cite{Chen2012,Lee2013,Flint2013,UdagawaMoessner2013,RauKee2014,Udagawa2012} and demonstrated that long-range spin-ice order reconstructs the Ir bands~\cite{Yao2018}. However, the real-space picture in the most relevant regime, where the Pr texture is correlated but lacks long-range order, remains completely unexplored.

We initially hypothesized that the magnetic frustration of the Pr spin moments could be used to control or destabilize the metal-insulator transition. To test this, we treated the Pr atoms as classical moments and evaluated the phase boundary. Our results show that the boundary does indeed shift, but the operative mechanism is specifically the local exchange field generated by the spin configuration. As the exchange coupling $J_K$ grows, the insulating gap closes by an amount directly dependent on the spin texture's monopole content. Consequently, materials poised near the boundary are highly sensitive to the local spin-ice physics taking place on the Pr lattice; wherever the spin texture varies in space, the local exchange field varies, directly modulating the insulating gap.

\section{The Model}

We begin with the iridium electrons. The basic kinetic energy is described by a nearest-neighbor hopping parameterized by $t$. Due to the large Ir nuclei, strong atomic spin-orbit coupling entangles the internal spin with the external crystal momentum, effectively locking the states into a $J_{\mathrm{eff}} = 1/2$ subspace. When an electron hops between local Ir axes, there is a finite amplitude for the pseudospin to flip. This spin-orbit hopping is tracked via right-hand rule book-keeping vectors $\nu_{ij}$~\cite{Kurita2011}. Putting this together, the non-interacting iridium lattice Hamiltonian is:
\begin{equation}
H_{Ir} = -t\sum_{\langle i, j\rangle} (c_{j}^{\dagger}c_{i}+h.c.) + i \lambda\sum_{\langle i, j\rangle} (\nu_{ij} \cdot c_{j,\sigma}^{\dagger}\tau_{\sigma,\sigma '} c_{i,\sigma '} + h.c.)
\end{equation}
The on-site electron-electron interaction $H_U = U\sum_{i} c_{i \uparrow}^{\dagger}c_{i \uparrow}c_{i \downarrow}^{\dagger}c_{i\downarrow}$ is made tractable via a mean-field approach. By neglecting quantum fluctuations and defining the site magnetization order parameter $\vec S_i$, the interaction decomposes to isolate the magnetic fields:
\begin{equation}
H_{U, MF} = (\vec S_i)^2 - 2 \vec S_i \cdot \hat J_i
\end{equation}
where $\hat J_i$ is the pseudospin operator. For moderately large values of $U$, this combined model drives the iridate sublattice into an insulating ground state. The interaction supplies the moments, and the spin-orbit coupling selects their all-in-all-out (AIAO) arrangement~\cite{Wan2011,Witczak-Krempa2012}. The eight-band Hartree-Fock model ($H_{Ir} + H_{U, MF}$) supplies this physics, and we take it over unchanged; the rare-earth coupling is what we add to it.

We now turn to the praseodymium ion. The two $Pr^{+3}$ electrons occupy a well-localized $4f^2$ orbital in which the atomic spin-orbit coupling and strong crystal field anisotropy render the ground state as a magnetic non-Kramer's doublet. 

The coupling itself is the f-d exchange,
\begin{equation}
H_{f-d} = J_{K} \sum_{\langle i, I\rangle} \vec S_{I} \cdot c_{i\alpha}^{\dagger} \vec \sigma_{\alpha\beta} c_{i\beta},
\end{equation}
where $\vec\sigma$ acts on the Ir spin index and $\vec S_I$ points into or out of a Pr-tetrahedra. Each Pr site thus imposes a static Zeeman field on its Ir neighbors along $\hat z_I$.

We neglect the coupling between the transverse components of $\vec S_I$ and the iridium electron density (which is the symmetry allowed coupling of these non-Kramers ``spins''\cite{Yao2018}). This means that our model does not contain the Kondo singlet, but it affords us the computation efficiency that comes with classical spins. This approximation is warranted as we are concerned here with the influence of the of praseodymium magnetism on the metal-to-insulator transition.

Kondo screening truncates the range of the RKKY interaction, and it is this truncation which justifies the effective near-neighbor Hamiltonian $H_{RKKY}$ below~\cite{Nakatsuji2006,Yao2018}. We note that this truncation is strictly an assumption of our calculation: it holds in the regions where the Kondo effect is strong, and the tunneling data of Ref.~\cite{Kavai2021} give direct evidence that the screening is not uniform across the surface. 

Our scheme is one-way. We sample praseodymium spin-ice configurations by classical Monte Carlo. Each configuration is frozen as a static Ising field, and the iridium Hartree-Fock mean field is solved in its presence. Freezing each configuration is a good approximation, as there is a large separation of time scales: neutron and muon data see the Pr fields as quasistatic over a wide temperature range~\cite{MacLaughlin2015,MacLaughlin2009}, and spin-ice monopole dynamics is slower still~\cite{Samarakoon2021,Hallen2022}, while the Ir electrons are metallic. On the time scale that sets the electronic structure the texture is quenched. Yao and Chen studied the Luttinger semimetal perturbed by the long-range Melko-Hertog-Gingras spin ice order, finding that the static spin texture reconstructs the Ir bands; this frames the U-$J_K$ calculation we undertake~\cite{Yao2018}. We depart from that work in the state imposed. Melko-Hertog-Gingras order is periodic, so its Ir problem is a band structure at a definite wavevector; we instead sample the frustrated manifold, with ice correlations but no long-range order, and must solve the Ir problem in real space configuration by configuration. What it returns is a distribution of local environments, which lets us ask, as Ref.~\cite{Yao2018} could not, whether a single sample fractures into regions of differing metallicity.

\begin{figure*}
    \centering
    \includegraphics[width=0.85\linewidth]{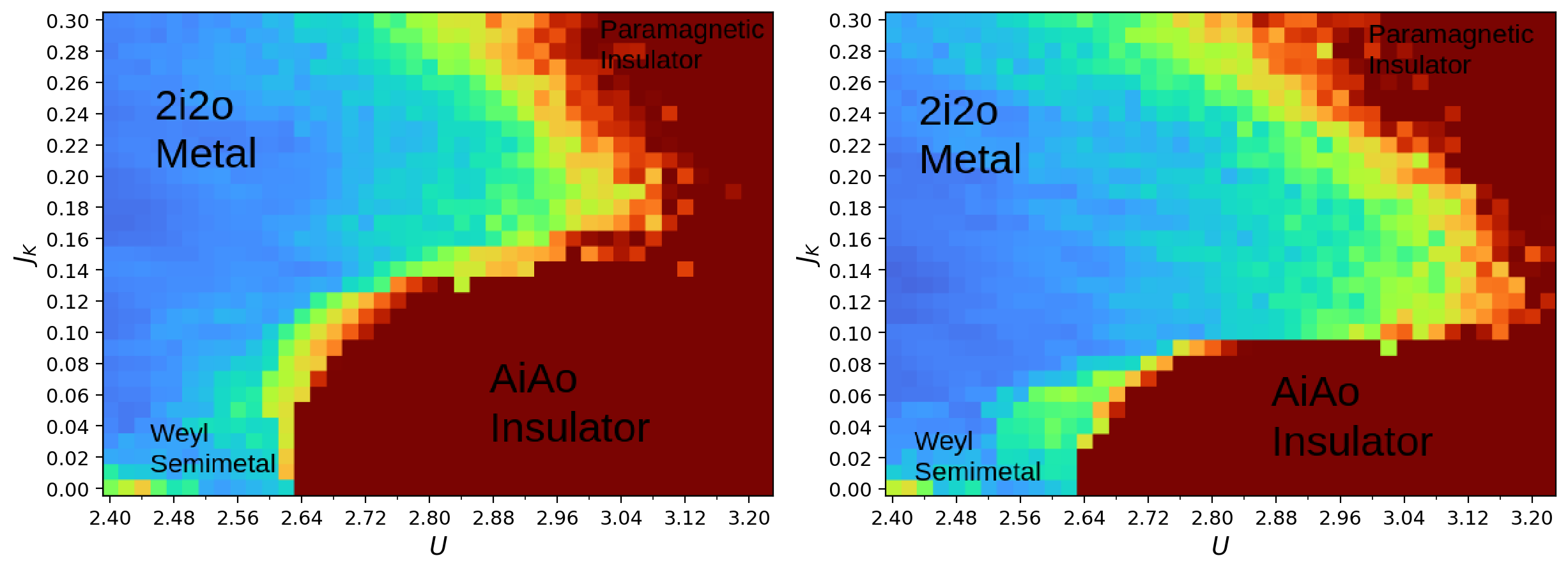}
    \caption{The U-$J_K$ phase diagram of the coupled Pr-Ir model. Figures: praseodymium texture (left: spin ice, $J=0$; right: jellyfish, $J=1/4$). Columns: band gap (left) and AIAO order parameter (right), computed from the iridium Hartree-Fock mean field over Monte Carlo-sampled Pr configurations at $T=1$ K on an $N=864$ cluster. The AIAO insulator occupies large $U$; increasing $J_K$ smoothly closes the gap.}
    \label{fig:phase}
\end{figure*}

Since the praseodymium moments are heavily screened, we define the effective Hamiltonian of the $J_1-J_2-J_3$ spin ice model as
\begin{equation}
\begin{split}
H_{RKKY} &= J_1\!\!\sum_{\langle IJ\rangle}\! \vec{S}_I\! \cdot\! \vec{S}_J + J_2\!\! \sum_{\langle\langle IJ\rangle\rangle}\! \vec{S}_I\! \cdot\! \vec{S}_J + J_3\!\! \sum_{\langle\langle\langle IJ\rangle\rangle\rangle}\! \vec{S}_I\! \cdot\! \vec{S}_J
\end{split}
\end{equation}
including the first, second and third nearest-neighbor coupling. We follow Udagawa et al. in parameterizing this model; choosing $J_1 = J_0$, $J_2 = J_{3a} = J_0 J$, and $J_{3b} = 0$, where $J_0$ is the overall scale and $J$ is the phase-line parameter among all-in-all-out, spin ice, and metastable like-charge clustering states. 

\section{Results: U-$J_K$ Phase Diagram}

Our chief result pertains to the U-$J_K$ phase diagram (Fig.~\ref{fig:phase}). Set beside the phase diagram assembled for $Nd_2Ir_2O_7$ from transport, neutron, and domain-wall evidence, our computed diagram reproduces the shared features: the AIAO insulator at large $U$ and the narrow Weyl window at its boundary~\cite{Tomiyasu2012,Tian2015,Ma2015}. As the two materials are nearly the same system up to the rare-earth physics, this lends credence to the model.

\begin{figure}
    \centering
    \includegraphics[width=\linewidth]{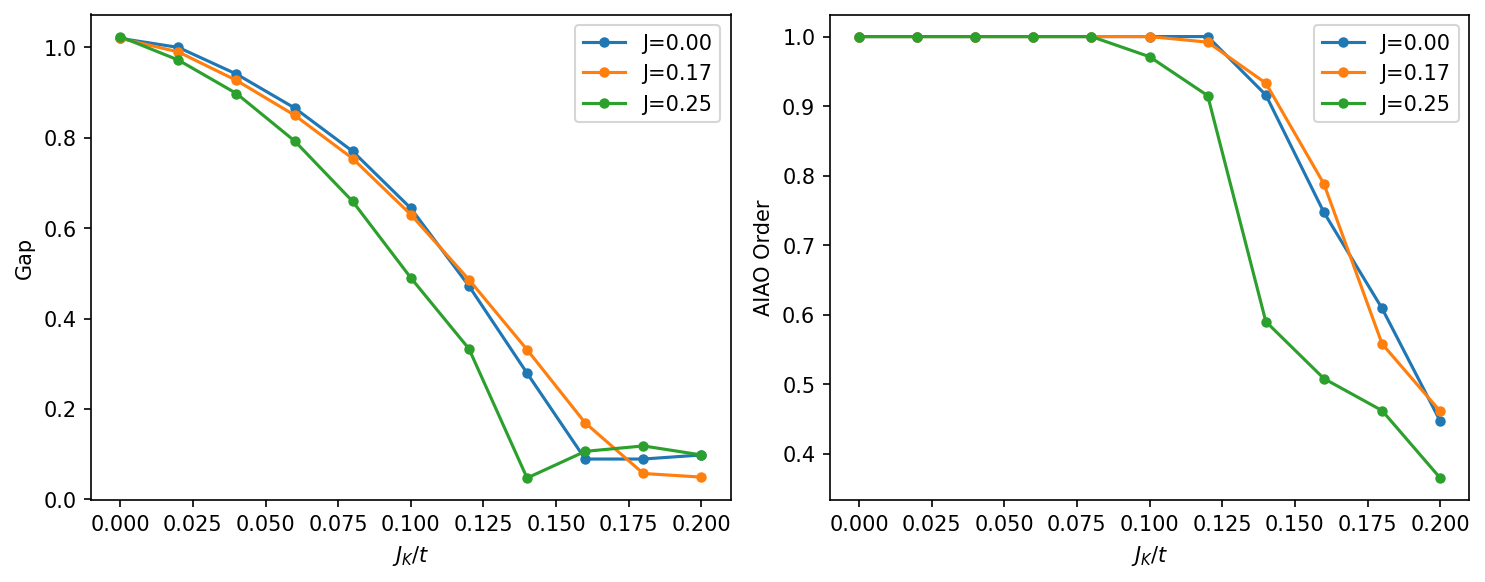}
    \caption{Gap (a) and AIAO order (b) versus $J_K$ for Pr textures along the ice--jellyfish line. The critical coupling shifts by $\Delta J_{Kc}\approx 0.03\,t$ between endpoints.}
    \label{fig:jline}
\end{figure}

From the cut along $J_K$ (Fig. \ref{fig:jline}), we see that increased coupling to the rare-earth moments smoothly closes the gap. The AIAO order deteriorates shortly after the closing of the gap, all while the moment length remains untouched. This is in contrast with the sharp transition that occurs at the interaction-driven critical point $U_c$. This means that, at the level of this model, any spatial fluctuations in the Kondo coupling result in smooth variation of the metallicity and the magnetic order.

Where does the material itself sit? Estimates place the Coulomb scale just below the insulating threshold, and the Kondo physics of the material demands a coupling strong enough to screen, but not so strong as to gap the spectrum outright. 

Sweeping the praseodymium texture along the ice-jellyfish phase line shifts the metal-insulator boundary in $J_K$ (Fig.~\ref{fig:jline}). The difference between ice and jellyfish is $\Delta J_{Kc} \approx 0.03\,t$. If $Pr_2Ir_2O_7$ sits anywhere near this boundary, these results suggest it will be highly susceptible to the local praseodymium magnetism. 

\section{Results: Large-Surface Local Density of States}

\begin{figure}
    \centering
    \includegraphics[width=0.85\linewidth]{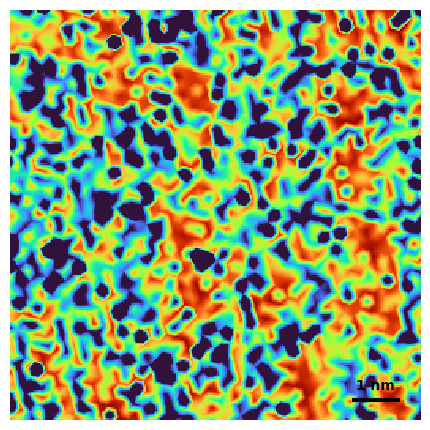}
    \caption{Synthetic local density of states over a 10 nm praseodymium field at $T=1$ K, $J=1/4$, $U=2.8$, $J_K=0.11$. Insulating islands appear within a metallic sea. The praseodymium system contains 24{,}576 spins and the iridium system contains 864 sites.}
    \label{fig:ldos}
\end{figure}

Given the sensitivity of the phase diagram, local spin textures should naturally induce spatial fluctuations in metallicity, which we can illustrate synthetically. While Monte Carlo allows us to simulate large spin systems (on the order of 10 nm), computing full band structures at this scale is computationally intractable. Instead, we scan a small iridium model cluster over the large praseodymium-spin surface. By computing the local density of states at each point we derive an object analogous to an STM measurement.

At each step in convolution, a moving window of praseodymium spins is taken as input for the MFT. The result is a set of iridium tight-binding models which vary with the underlying Ising field. By computing the appropriate Green's function at each point we arrive at a spatially rich simulation of STM. Within the Tersoff-Hamann picture the low-bias tunneling conductance is proportional to the local density of states~\cite{TersoffHamann1985},
\begin{equation}
\rho(\mathbf{r},\omega) = -\frac{1}{\pi}\sum_{a}\mathrm{Im}\,G^{R}_{aa}(\mathbf{r},\mathbf{r};\omega)
\end{equation}
\begin{equation}
G^{R}(\omega)=\big[(\omega+i\epsilon)\,\mathds{1}- H_{\mathrm{MF}}\big]^{-1}
\end{equation}
where $a$ runs over the orbital and spin indices at position $\mathbf{r}$ and $H_{\mathrm{MF}}$ is the eight-band Hartree-Fock Hamiltonian of the local cluster; the map records the Fermi-level value, $dI/dV(\mathbf{r},eV)\propto\rho(\mathbf{r},eV)$.
Convolving the mean-field theory over the Monte Carlo textures yields synthetic tunneling maps (Fig.~\ref{fig:ldos}), and these maps bear a qualitative resemblance to the measured conductance maps of Ref.~\cite{Kavai2021}: distinct spectral regions, arranged as islands within a sea in a self-similar pattern.

Because $T_K \propto W e^{-1/2\rho J_K}$ depends exponentially on $\rho(E_F)$ while $T_{RKKY} \propto \rho J_K^2$ depends only algebraically, a modest, texture-induced variation in $\rho(E_F)$, of the magnitude our maps display, produces a large spatial variation in $T_K$, enough to push some regions to the Kondo-screened side ($T_K > T_{RKKY}$) and others to the Kondo-destroyed, magnetically-correlated side. The frustrated Pr spin texture thus need not gap the spectrum outright to fracture the surface; it need only tilt the local Doniach balance. In this light the computed $\rho(E_F)$ landscape is not a literal map of the tunneling phases but the input that selects them; a frustration-driven mechanism for the coexistence of Kondo and Kondo-destroyed regions~\cite{Kavai2021}.

\section{Discussion}
While our classically simulable model successfully captures a frustration-driven route to electronic inhomogeneity seen in STM\cite{Kavai2021}, they are unlikely to be the direct cause of the inhomogenaity. The temperature at which our textures are prepared is an order of magnitude lower than the temperatures at which the experimental inhomogeneity is observed, and the spatial scale of our islands is smaller than the measured domains also by an order of magnitude. Further, our results would implicity assume the spins are static on the timescales of the experiment which is unlikely. However, the geometry is the right one, the experiments show self-similarity, they do not need structural disorder of any kind and complements the cleaving-damage scenario~\cite{Song2025} as the two mechanisms make different predictions for how the inhomogeneity should respond to field and thermal history. 

It would be interesting to study the anomalous Hall response. Udagawa et al.\ identified monopole textures as time-reversal-breaking objects generating the anomalous Hall response~\cite{Udagawa16,Machida2007,Machida2010}; if the textures under our insulating islands keep that character and are as static as we assume, the islands should carry a local anomalous Hall contrast against the metal. The step from individual jellyfish to our extended regions is not established, though it would need a chirality-resolved analysis of the sampled textures, which we have not done. 

Our results demonstrate a sensitivity of the metal-to-insulator transition in the pyrochlore iridates to the behavior of local rare-earth moments. They link the inhomogenaity in STM and the metallicity of Pr$_2$Ir$_2$O$_7$ to the underlying rare-earth magnetism. 

\begin{acknowledgments}
This work was supported in part by the Office of Naval Research grant number N00014-15-1-2760.
\end{acknowledgments}

\bibliographystyle{apsrev4-2}
\bibliography{references}

\end{document}